\documentclass[prl,aps,twocolumn,showpacs]{revtex4}
\input{psfig.sty}

\newcommand{\la}{\left\langle}
\newcommand{\ra}{\right\rangle}
\newcommand{\EPL}{{\it Europhys.~Lett.~}}
\newcommand{\PRL}{{\it Phys.~Rev.~Lett.~}}
\newcommand{\PR}{{\it Phys.~Rev.~}}
\newcommand{\JCP}{{\it J.~Chem.~Phys.~}}
\newcommand{\JPCM}{{\it J.~Phys.: Condens.~Matter~}}
\newcommand{\MP}{{\it Mol.~Phys.~}}

\newcommand{\EPJ}{{\it Eur.~Phys.~J.~}}

\begin{document}

\author{A. R. Denton}
\affiliation
{Department of Physics, North Dakota State University, Fargo, ND, 58105-5566}

\title{Nonlinear Screening and Effective Interactions
in Charged Colloids}

\date{\today}
\begin{abstract}
Influences of nonlinear screening on effective interactions 
between spherical macroions in charged colloids are described 
via response theory.
Nonlinear screening, in addition to generating effective many-body
interactions, is shown to entail essential corrections to the pair
potential and one-body volume energy. Numerical results
demonstrate that nonlinear effects can substantially modify
effective interactions and thermodynamics of aqueous, low-salt
suspensions of highly-charged macroions.
\end{abstract}

\pacs{PACS numbers: 82.70.Dd, 82.45.-h, 05.20.Jj}

\maketitle



Colloidal particles, typically nanometers to microns in size,
commonly acquire electric charge in solution~\cite{Schmitz}.
Practical examples are latex microspheres in paints, ionic micelles
in detergents, and platelets in clay suspensions.  Beyond 
traditional applications, {\it e.g.}, in the chemical, pharmaceutical, 
and mining industries, the rich thermal and optical properties 
of charged colloidal crystals have inspired recent proposals 
for novel technologies, such as fast optical switches and
photonic band-gap materials~\cite{tech}.

Among the various interparticle interactions at play in charged
colloids, electrostatic interactions strongly affect
thermodynamic stability and materials properties~\cite{CC}. 
The classic theory of Derjaguin, Landau, Verwey, and Overbeek
(DLVO)~\cite{DLVO} describes the bare Coulomb interactions
between pairs of macroions as {\it screened} by 
surrounding microions (counterions and salt ions) in solution. 
In recent years, the general validity of the DLVO screened-Coulomb 
pair potential has been questioned in light of reported
observations of phase separation~\cite{Ise} and related
anomalies~\cite{Grier} in strongly deionized suspensions.

The DLVO pair potential was originally derived by linearizing the
mean-field Poisson-Boltzmann (PB) equation for the electrostatic
potential around a macroion~\cite{DLVO}.
Density-functional~\cite{vRH} and 
linear response~\cite{Silbert,Denton} theories 
offer alternative routes to
effective interactions, but also yield, as an important
by-product, a one-body (volume) energy that contributes to the
total free energy. Common to all approaches is the simplifying
assumption that the microion densities respond {\it linearly} to
the macroion charge density. While linearization is often a
reasonable approximation, its justification is uncertain in cases
of high macroion charge and low salt concentration (ionic
strength), precisely those conditions under which anomalous phase
behavior is observed.

The purpose of this paper is to define the range of validity of
the linearization approximation by extending the DLVO theory
beyond linear screening. Applying response theory to the primitive
model of charged colloids, leading-order nonlinear corrections to
the effective pair potential and volume energy are obtained and
shown to qualitatively affect thermodynamics of concentrated, 
deionized suspensions of highly charged macroions.

The model system comprises $N_m$ spherical macroions, of hard-core
diameter $\sigma$ and uniformly-distributed surface charge $-Ze$
($e$ being the elementary charge), and $N_c$ point counterions of
charge $ze$ dispersed in a symmetric electrolyte in a volume $V$
at temperature $T$.  The electrolyte consists of $N_s$ point salt
ions of charge $ze$ and $N_s$ of charge $-ze$ in a uniform
solvent, characterized entirely by a dielectric constant
$\epsilon$. The microions thus number $N_+=N_c+N_s$ positive and
$N_-=N_s$ negative, for a total of $N_{\mu}=N_c+2N_s$. Global
charge neutrality constrains macroion and counterion numbers via
$ZN_m=zN_c$.

Initially ignoring salt, the Hamiltonian decomposes into three
terms: $H=H_m+H_c+H_{mc}$.  Here $H_m$ is the bare macroion
Hamiltonian involving Coulomb interactions,
$v_{mm}(r)=Z^2e^2/\epsilon r$, between macroion pairs at
center-center separation $r>\sigma$; $H_c$ is the counterion
Hamiltonian involving Coulomb pair interactions,
$v_{cc}(r)=z^2e^2/\epsilon r$; and $H_{mc}$ is a sum over
macroion-counterion pair interactions of the form
\begin{equation}
v_{mc}(r)~=~\left\{ \begin{array} {l@{\quad\quad}l}
\frac{\displaystyle -Zze^2}{\displaystyle \epsilon r}, & r>\sigma/2 \\
\frac{\displaystyle -Zze^2}{\displaystyle \epsilon\sigma/2}\alpha,
& r<\sigma/2.
\end{array} \right.
\label{vmcr}
\end{equation}
As the interaction for $r<\sigma/2$ is arbitrary, $\alpha$ is
chosen to minimize counterion penetration inside the
core~\cite{vRH}.

The mixture of macroions and counterions is formally reduced to an
equivalent one-component system of ``pseudo-macroions" by
performing a restricted trace over counterion coordinates, keeping
the macroions fixed. Denoting counterion and macroion traces by
$\la~\ra_c$ and $\la~\ra_m$, respectively, the canonical partition
function is
\begin{equation}
{\cal Z}~=~\la\la\exp(-\beta H)\ra_c\ra_m
~=~\la\exp(-\beta H_{eff})\ra_m, \label{part}
\end{equation}
where $H_{eff}=H_m+F_c$ is the effective one-component
Hamiltonian, $\beta=1/k_BT$, and
\begin{equation}
F_c~=~-k_BT\ln\la\exp\Bigl[-\beta(H_c+H_{mc})\Bigr]\ra_c
\label{Fc1}
\end{equation}
is the free energy of a nonuniform gas of counterions in the
presence of the macroions. Adding to $H_c$ (and subtracting from
$H_{mc}$) the energy of a uniform compensating negative
background, $E_b$, converts $H_c$ to the Hamiltonian of a
classical one-component plasma (OCP) of counterions. In practice,
the OCP is weakly correlated, with coupling parameter
$\Gamma=\lambda_B/a_c \ll 1$, where $\lambda_B=\beta
e^2/\epsilon$ is the Bjerrum length and $a_c=(3/4\pi
n_c)^{1/3}$.  The OCP has mean density $n_c=N_c/V(1-\eta)$, since
the counterions are excluded (with the background) from the
macroion core volume fraction,
$\eta=\frac{\pi}{6}(N_m/V)\sigma^3$.

The counterion free energy is approximated
by regarding the macroions as an ``external" potential for the OCP
and invoking perturbation theory~\cite{Silbert,HM}:
\begin{equation}
F_c~=~F_{OCP}~+~\int_0^1{\rm d}\lambda\,\la H'_{mc}\ra_{\lambda},
\label{Fc2}
\end{equation}
where $F_{OCP}=-k_BT\ln\la\exp[-\beta(H_c+E_b)]\ra_c$ is the OCP
free energy, $H'_{mc}=H_{mc}-E_b$, and the $\lambda$-integral
charges the macroions. In terms of Fourier components,
\begin{equation}
\la H_{mc}\ra_{\lambda}~=~\frac{1}{V}\sum_{\bf k} \hat v_{mc}(k)
\hat\rho_m({\bf k}) \la\hat\rho_c(-{\bf k})\ra_{\lambda},
\label{Hmc2}
\end{equation}
where $\hat v_{mc}(k)$
is the Fourier transform of Eq.~(\ref{vmcr}) and $\hat\rho_m({\bf
k})$ and $\hat\rho_c({\bf k})$ are Fourier components of the
macroion and counterion number densities.
Now, to second order in the macroion density, we may write (for
$k\neq 0$)
\begin{eqnarray}
\hat\rho_c({\bf k})=&\chi&^{(1)}(k) \hat v_{mc}(k)\hat\rho_m({\bf
k})+\frac{1}{V}\sum_{{\bf k}'}
\chi^{(2)}({\bf k}',{\bf k}-{\bf k}') \nonumber \\
&\times&\hat v_{mc}(k') \hat v_{mc}(|{\bf k}-{\bf k}'|)
\hat\rho_m({\bf k}') \hat\rho_m({\bf k}-{\bf k}')
\label{rhock-lambda}
\end{eqnarray}
where $\chi^{(1)}$ and $\chi^{(2)}$ are, respectively, the linear
and first nonlinear response functions of the OCP~\cite{Louis}.

To specify the response functions, we adopt the random phase
approximation (RPA), which equates the two-particle direct
correlation function of the OCP to its exact asymptotic limit:
$c^{(2)}(r)=-\beta v_{cc}(r)$~\cite{HM}. 
Within the RPA,
$\chi^{(1)}$ and $\chi^{(2)}$ have analytic forms
\begin{equation}
\chi^{(1)}(k)~=~\frac{-\beta n_c}{1+\kappa^2/k^2}, \label{chi1}
\end{equation}
where $\kappa=\sqrt{4\pi n_cz^2\lambda_B}$, and
\begin{equation}
\chi^{(2)}=\frac{-\beta^2n_c/2}
{(1+\kappa^2/k^2)(1+\kappa^2/k'^2) (1+\kappa^2/|{\bf k}+{\bf
k}'|^2)}~. \label{chi2}
\end{equation}

With the response functions specified, the counterion density can
be explicitly determined from Eq.~(\ref{rhock-lambda}).
Penetration inside the macroion cores is reduced, to at most a few
per cent, by choosing $\alpha=\kappa\sigma/(2+\kappa\sigma)$ in
Eq.~(\ref{vmcr}). Salt is incorporated via additional response
functions for the multiple microion species.  In the process,
$\kappa$ is modified by replacing the counterion density, $n_c$,
by the total microion density, $n_{\mu}=n_++n_-$, where
$n_\pm=N_\pm/V(1-\eta)$ are the densities of positive and negative
microions.

Combining Eqs.~(\ref{Fc2})-(\ref{chi2}), the effective Hamiltonian
can be written as a sum of one-, two-, and three-body terms. The
one-body term, or volume energy, is the sum of
\begin{equation}
E_0=F_{OCP}-\frac{k_BTN_m}{2}\left(Z^2\lambda_B
\frac{\kappa}{1+\kappa\sigma/2}
-Z\frac{n_c}{n_{\mu}}\right), \label{E03}
\end{equation}
which is the linear response volume energy~\cite{Denton}, and
\begin{equation}
\Delta E=k_BT\frac{N_m}{6}\frac{n_c}{n_{\mu}^2}\int{\rm d}{\bf
r}\,\left(\left[\rho_0( r)\right]^2-\frac{1}{n_{\mu}}
\left[\rho_0(r)\right]^3\right) \label{Delta-E1r-salt}
\end{equation}
which is the first nonlinear correction, with
\begin{equation}
\rho_0(r)=\frac{Z}{z}~\frac{\kappa^2}{4\pi}
\left(\frac{e^{\kappa\sigma/2}}{1+\kappa\sigma/2}\right)
\frac{e^{-\kappa r}}{r}, \qquad r>\sigma/2 \label{rho0r}
\end{equation}
being the linear response counterion density orbital around a
single macroion.  In Eq.~(\ref{rho0r}), $\kappa$ can be identified
now as an inverse Debye screening length.
The first and second terms on the right side of Eq.~(\ref{E03})
account, respectively, for the counterion entropy and the
macroion-counterion electrostatic interaction energy, while
Eq.~(\ref{Delta-E1r-salt}) corrects the latter term for nonlinear
screening.

The two- and three-body terms in $H_{eff}$ involve effective pair
and triplet interactions. The pair interaction, $v^{(2)}(r)$, is
the sum of the linear response form~\cite{Denton}
\begin{equation}
v^{(2)}_0(r)=\frac{Z^2e^2}{\epsilon}\left(\frac{e^{\kappa\sigma/2}}
{1+\kappa\sigma/2}\right)^2\frac{e^{-\kappa r}}{r}, \qquad
r>\sigma, \label{v20r}
\end{equation}
{\it i.e.}, the DLVO potential~\cite{DLVO} (aside from the
excluded-volume adjustment to $\kappa$), and the nonlinear
correction
\begin{eqnarray}
\Delta v^{(2)}(r)=&-&k_BT\frac{n_c}{n_{\mu}^3}\int{\rm d}{\bf
r}'\,\rho_0(|{\bf r}-{\bf r}'|)\rho_0(r') \nonumber \\
 &\times&\left[\rho_0(r') -\frac{n_{\mu}}{3}\right].
\label{Delta-v2effr-salt}
\end{eqnarray}
Finally, the triplet interaction is
\begin{eqnarray}
v^{(3)}({\bf r}_1,{\bf r}_2,{\bf r}_3)
=&-&k_BT\frac{n_c}{n_{\mu}^3}\int{\rm d}{\bf r}\,\rho_0(|{\bf
r}_1-{\bf r}|) \nonumber \\
 &\times&\rho_0(|{\bf r}_2-{\bf r}|)
\rho_0(|{\bf r}_3-{\bf r}|). \label{v3effr-salt}
\end{eqnarray}
Note that, with increasing salt concentration or decreasing
macroion volume fraction ($n_c/n_{\mu}\to 0$), the nonlinear
corrections, $\Delta E$ and $\Delta v^{(2)}(r)$, and the effective
triplet interaction, all diminish in relative magnitudes. Thus,
nonlinear effects are strongest for deionized suspensions.

Equations (\ref{Delta-E1r-salt}), (\ref{Delta-v2effr-salt}), and
(\ref{v3effr-salt}) are the main theoretical results of the paper.
Equation (\ref{Delta-v2effr-salt}), which is a {\it bulk}
nonlinear correction to $v_0^{(2)}(r)$, is similar in structure to
a {\it wall-induced} pair interaction derived in
Ref.~\cite{Goulding} [Eq.~(13) therein]. Furthermore,
Eq.~(\ref{v3effr-salt}) is consistent with the triplet force
derived in Ref.~\cite{LA} [Eq.~(33) therein]. The response theory,
in neglecting short-range correlations between microions, is
formally equivalent to the mean-field PB theory. Its scope is 
therefore limited to long-range interactions between macroions. 
However, whereas numerical solutions of the nonlinear PB equation 
are practical only within symmetric Wigner-Seitz cells, 
characteristic of crystals (PB cell models)~\cite{CC,Alexander},
the effective interactions derived here apply, in principle, to
any bulk phase.

We can now quantify the practical significance of nonlinear 
screening by evaluating the effective pair interaction and 
volume energy for selected system parameters. All results 
presented are for bulk, aqueous suspensions ($\lambda_B=0.714$ nm) 
and monovalent counterions ($z=1$). Analytic reduction of
Eqs.~(\ref{Delta-E1r-salt}) and (\ref{Delta-v2effr-salt}) was
checked by Monte Carlo integration. Figure 1 shows the effective
pair potential for macroions of diameter $\sigma=100$ nm, valence
$Z=700$, and volume fraction $\eta=0.01$, at ionic strength
$c_s=1~\mu$M.  The chosen charge is comparable to the maximum
renormalized charge (allowing for counterion condensation),
$Z^*=C\sigma/\lambda_B$, where $C=O(10)$~\cite{Alexander}.
\unitlength1mm
\begin{figure}
\smallskip
\begin{center}
\begin{picture}(80,60)
\put(0,0){\psfig{figure=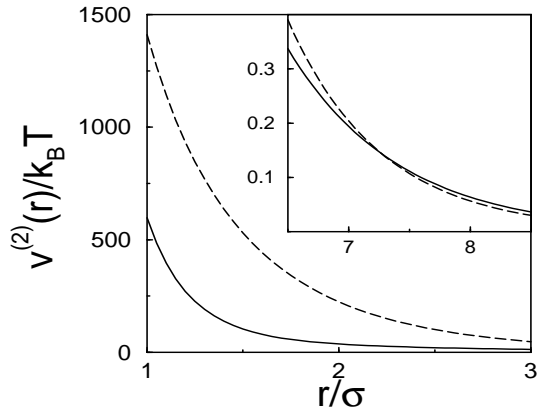,width=80mm,height=60mm}}
\end{picture}
\end{center}
\caption[]{Effective pair interaction between macroions of
diameter $\sigma=100$ nm, valence $Z=700$, and volume fraction
$\eta=0.01$ at ionic strength $c_s=1~\mu$M.  Solid curve:
nonlinear screening. Dashed curve: linear screening (DLVO). Inset:
cross-over from $e^{-\kappa r}/r$ to $e^{-\kappa r}$ behavior at
longer range.} \label{Fig1}
\end{figure}
At short and intermediate ranges, nonlinear screening sharpens the
counterion profile around a macroion, which, as seen in Fig.~1,
weakens the pair repulsion. This result is consistent, if only
qualitatively, with bulk measurements~\cite{Franck} of pair
interactions shorter in range than predicted by DLVO theory. At
longer range (inset), nonlinear screening changes the asymptotic
form to $v^{(2)}(r)\sim e^{-\kappa r}$, a qualitative departure
from the more rapidly decaying DLVO form, $v^{(2)}_0(r)\sim
e^{-\kappa r}/r$.
As the macroion charge increases, higher-order nonlinear effects 
ultimately manifest themselves.  In fact, for values of $Z>Z^*$
the effective pair potential can actually become attractive. 
Since pair attractions are mathematically excluded within 
the PB theory~\cite{Neu}, higher-order nonlinear terms
must then also contribute in such extreme cases. 
Within physical parameter ranges ($Z<Z^*$), however, 
$v^{(2)}(r)$ is always repulsive.
Interestingly, the short-to-intermediate-range part 
of the nonlinear pair potential may still be reasonably fit 
by a DLVO-like potential.

The cleanest test of the theory is comparison with 
{\it ab initio} simulation~\cite{Tehver}, which,
like our response theory, ignores short-range counterion 
correlations.
Figure~2 presents the comparison for the total potential energy of
interaction between a pair of macroions, of diameter $\sigma=106$
nm and valence $Z=200$, in a cubic box of length $530$ nm with
periodic boundary conditions (taking into account image
interactions) in the absence of salt.  The agreement is
essentially perfect, albeit for a case of relatively weak
nonlinearity. Further simulations of more highly charged macroions
are needed to more severely test the theory.
\unitlength1mm
\begin{figure}
\begin{center}
\begin{picture}(80,60)
\put(0,0){\psfig{figure=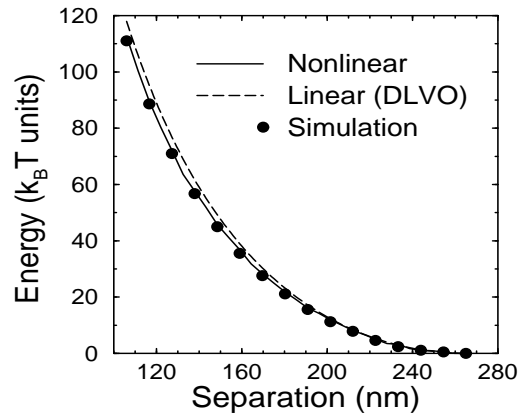,width=80mm,height=60mm}}
\end{picture}
\end{center}
\caption[]{Total interaction potential energy for two macroions
of diameter $\sigma=106$ nm and valence $Z=200$, at zero ionic
strength, in a cubic box of length 530 nm with periodic
boundary conditions. Solid curve: nonlinear screening prediction.
Dashed curve: linear screening (DLVO) prediction.
Symbols: {\it ab initio} simulation data~\cite{Tehver}.} \label{Fig2}
\end{figure}
A measure of the impact of nonlinear screening on thermodynamics
is the relative magnitude of $\Delta v^{(2)}(r)$ at the mean
nearest-neighbor separation, $r_{nn}$, where $v^{(2)}(r)$ makes its
dominant contribution to the potential energy. Figure~3 maps out,
in the space of volume fraction and ionic strength, the boundary
of the region within which $|\Delta v^{(2)}(r_{nn})|$ exceeds
$v^{(2)}_0(r_{nn})$ by at least 20\% for the fcc crystal
structure: $r_{nn}/\sigma=2^{-1/2}(2\pi/3\eta)^{1/3}$. With
increasing $Z$ and decreasing $c_s$, the threshold $\eta$ {\it
decreases}.

Thermodynamic properties of charged colloids are influenced also
by the volume energy, $E$, which contributes to the free energy, $F$,
and pressure, $P=-(\partial F/\partial V)_T$. The density
dependence of $E$ has been predicted -- within
linear screening -- to drive an instability toward phase
separation at low ionic strengths~\cite{vRH,Warren-Chan}.  The
impact of nonlinearity is illustrated in Fig.~4, which plots the
volume energy contribution, $P_v=-(\partial E/\partial V)_T$, to
the total equation of state for a system characterized by
$\sigma=500$ nm, $Z=4000$, and $c_s=1~\mu$M. 

\unitlength1mm
\begin{figure}
\begin{center}
\begin{picture}(80,60)
\put(0,0){\psfig{figure=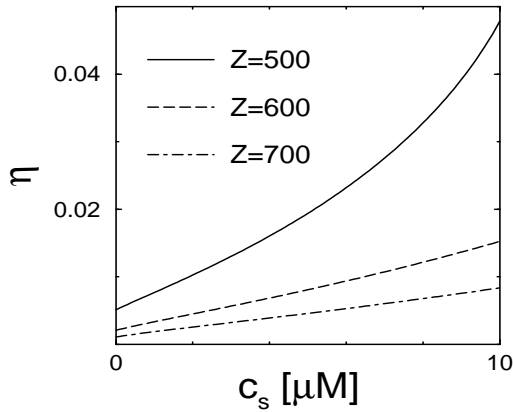,width=80mm,height=60mm}}
\end{picture}
\end{center}
\caption[]{Nonlinear deviations from DLVO theory for macroion
diameter $\sigma=100$ nm and valences, from top to bottom, $Z=500,
600, 700$. Systems with macroion volume fraction, $\eta$, and 
ionic strength, $c_s$, above the curve deviate from the DLVO 
pair potential by at least 20\% at the fcc-crystal 
nearest-neighbor distance.} \label{Fig3}
\end{figure}

\unitlength1mm
\begin{figure}
\begin{center}
\begin{picture}(80,60)
\put(0,0){\psfig{figure=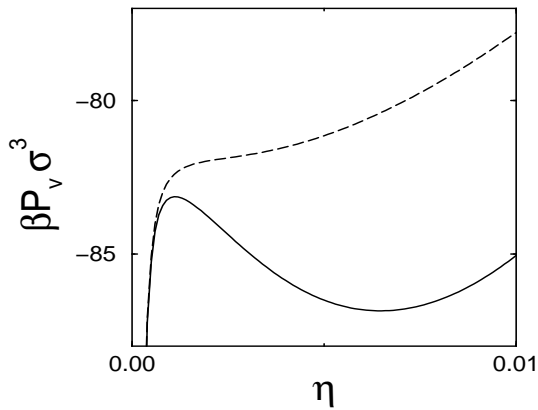,width=80mm,height=60mm}}
\end{picture}
\end{center}
\caption[]{Volume energy contribution, $P_v$, to the total reduced
pressure vs. volume fraction, $\eta$, for macroion diameter
$\sigma=500$ nm and valence $Z=4000$ at ionic strength
$c_s=1~\mu$M.  Solid curve: nonlinear screening prediction.  
Dashed curve: linear screening (DLVO) prediction.} \label{Fig4}
\end{figure}
Differences between the linear and nonlinear predictions for $P_v$
grow with increasing $Z$ and decreasing $c_s$, demonstrating the
capacity of nonlinear screening to significantly affect phase
stability of highly-charged, deionized suspensions. In fact, the
parameter ranges in which instabilities toward phase separation
have thus far been predicted~\cite{vRH,Warren-Chan} lie well
within the nonlinear regime. Future work will further explore
influences of nonlinearity, including the role of attractive
triplet interactions, on thermodynamics of charged colloids.

In summary, applying response theory to the primitive model of
charged colloids, we have consistently extended the
DLVO linear-screening theory by including leading-order
nonlinear contributions to effective interactions between
macroions. A major conclusion is that nonlinear corrections, aside
from generating effective many-body interactions, can
significantly modify the effective pair interaction and volume
energy, and thus thermodynamics, of deionized suspensions of
highly charged macroions.
While mean-field pair interactions are repulsive at long range,
nonlinear screening weakens shorter-range pair repulsion and 
generates effective three-body attractions. Nonlinear effects 
may thus facilitate any fluctuation-mediated short-range 
attractions~\cite{Holm} and help to explain
anomalous phase behavior observed in experiments~\cite{Ise,Grier}, 
as well as in recent simulations~\cite{Linse}.

Discussions with C.~N.~Likos, H.~L\"owen, and K.~S.~Schmitz, and
support of the Natural Sciences and Engineering Research Council
of Canada (Grant No.~217232-1999), are gratefully acknowledged.







\vspace{-0.5cm}
\begin{references}
\vspace{-0.5cm}
\bibitem{Schmitz}
K.~S.~Schmitz, {\it Macroions in Solution and Colloidal Suspension}
(VCH, New York, 1993).

\bibitem{tech}
G.~Pan, R.~Kesavamoorthy, and S.~A.~Asher, \PRL {\bf 78}, 3860 (1997);
J.~E.~G.~J.~Wijnhoven and W.~L.~Vos, {\it Science} {\bf 281}, 802 (1998).

\bibitem{CC}
For recent reviews, see J.-P.~Hansen and H.~L\"owen, 
{\it Ann.~Rev.~Phys.~Chem.} {\bf 51}, 209 (2000); 
L.~Belloni, \JPCM {\bf 12}, R549 (2000).

\bibitem{DLVO}
E.~J.~W.~Verwey, J.~T.~G.~Overbeek, {\it Theory of the Stability of
Lyophobic Colloids} (Elsevier, Amsterdam, 1948).

\bibitem{Ise}
K.~Ito, H.~Yoshida, and N.~Ise, {\it Science} {\bf 263}, 66 (1994).

\bibitem{Grier}
A.~E.~Larsen and D.~G.~Grier, {\it Nature} {\bf 385}, 230 (1997).


\bibitem{vRH}
R.~van Roij and J.-P.~Hansen, \PRL {\bf 79}, 3082 (1997); R.~van
Roij, M.~Dijkstra, and J.-P.~Hansen, \PR E {\bf 59}, 2010 (1999).

\bibitem{Silbert}
M.~J.~Grimson and M.~Silbert, \MP {\bf 74}, 397 (1991).

\bibitem{Denton}
A.~R.~Denton, \JPCM {\bf 11}, 10061 (1999); \PR E {\bf 62}, 3855
(2000).

\bibitem{HM}
J.-P.~Hansen and I.~R.~McDonald, {\it Theory of Simple Liquids},
$2 ^{nd}$ edition (Academic, London, 1986).

\bibitem{Louis}
For a recent application of nonlinear response theory 
to electronic systems, see A.~A.~Louis and N.~W.~Ashcroft, 
\PRL {\bf 81}, 4456 (1998).

\bibitem{Goulding}
D.~Goulding and J.-P.~Hansen, {\EPL} {\bf 46}, 407 (1999).

\bibitem{LA}
H.~L\"owen and E.~Allahyarov, \JPCM {\bf 10}, 4147 (1998).

\bibitem{Alexander}
S.~Alexander, P.~M.~Chaikin, P.~Grant, G.~J.~Morales, and P. Pincus,
\JCP {\bf 80}, 5776 (1984).

\bibitem{Franck}
R.~V.~Durand, C.~Franck, \PR E {\bf 61}, 6922 (2000).

\bibitem{Neu}
J.~C.~Neu, \PRL {\bf 82}, 1072 (1999);
J.~E.~Sader and D.~Y.~C.~Chan, 
{\it Langmuir} {\bf 16}, 324 (2000).

\bibitem{Tehver}
R.~Tehver, F.~Ancilotto, F.~Toigo, J.~Koplik, and J.~R.~Banavar,
\PR E {\bf 59}, R1335 (1999).

\bibitem{Warren-Chan}
P.~B.~Warren, \JCP {\bf 112}, 4683 (2000);
D.~Y.~C.~Chan, P.~Linse, and S.~N.~Petris, {\it Langmuir} 
{\bf 17}, 4202 (2001).

\bibitem{Holm}
R.~Messina, C.~Holm, K.~Kremer, \PRL {\bf 85}, 872 (2000); \EPJ E
{\bf 4}, 363 (2001); E.~Allahyarov, I.~D'Amico, and H.~L\"owen,
\PRL {\bf 81}, 1334 (1998).

\bibitem{Linse}
P.~Linse, \JCP {\bf 113}, 4359 (2000);
P.~Linse and V.~Lobaskin, \PRL {\bf 83}, 4208 (1999).




\end{references}
\end{document}